\documentclass{aa} 
\usepackage{natbib} 
\bibpunct{(}{)}{;}{a}{}{,} 
\usepackage{graphicx}
\defcitealias{perci2}{PSK03}  
\begin{document}
   \title{The distance to the Pleiades}

   \subtitle{Main sequence fitting in the near infrared}

   \author{S.M. Percival
          \inst{1}
          \and
          M. Salaris\inst{1}
          \and
          M.A.T. Groenewegen\inst{2}
          }

   \offprints{S.M. Percival}

   \institute{Astrophysics Research Institute,Liverpool John Moores 
University,
              Twelve Quays House, Egerton Wharf, Birkenhead, CH41 1LD, UK\\
              \email{smp,ms@astro.livjm.ac.uk}
         \and
             Instituut voor Sterrenkunde, PACS-ICC, Celestijnenlaan 200B,
              B-3001 Leuven, Belgium\\
             \email{groen@ster.kuleuven.ac.be}
             }

   \date{Received; accepted}

   \abstract{
Hipparcos parallax measurements of stars in the Pleiades notoriously result in
a cluster distance of 118 pc, which is approximately 10\% shorter than the 
`classical' result obtained from earlier main sequence (MS) fitting studies.
In an earlier paper we developed a purely empirical MS-fitting method 
in an attempt to address this problem.  This work produced 
conflicting results for the Pleiades between the $(B-V)$ and $(V-I)$ colour 
indices, indicating that the cluster's photometric metallicity is 
substantially lower than its generally accepted spectroscopic metallicity
of ${\rm [Fe/H]}=-0.03$.  We were able to reconcile the discrepancy by
assuming ${\rm [Fe/H]}=-0.4$, the appropriate metallicity indicated from
$(B-V)/(V-I)$ colour-colour plots, and the distance moduli obtained from the 
two colour indices were in agreement with the 
Hipparcos result, within the 1$\sigma$ errors.

With the release of the 2MASS All Sky Catalogue, we now apply our MS-fitting 
method to the Pleiades using the infrared colours in addition to the
optical bands, in order to test the plausibility of our earlier result.

Using the full field dwarf sample
and fitting in the $V/(V-K)$ and $K/(J-K)$ colour planes, we find 
that assuming a substantially subsolar metallicity does not produce distances 
in agreement with the $(B-V)$ and $(V-I)$ results.   However the infrared plus
$(V-I)$ distances are in mutual agreement when adopting the spectroscopic 
metallicity.  By considering only the field dwarfs with 
$M_V\leq 6$, i.e. brighter than the magnitude where the Pleiades $(B-V)$
colours start to be anomalous \citep{stauffer2003}, the infrared and
optical colour indices all yield consistent distances using the 
spectroscopic [Fe/H].
The concordant distances thus obtained 
from the $V/(B-V)$, $V/(V-I)$, $V/(V-K)$ and $K/(J-K)$ planes yield a mean of 
$133.8 \pm3$ pc, in excellent agreement with both the pre-Hipparcos 
MS-fitting results, and the most recent determinations from other methods.

We conclude that there are two distinct, and unrelated, issues affecting the 
Pleiades: 1) the Hipparcos parallax is in error by $\sim$ 10\%, as previously
claimed; 2) the $(B-V)$ colours of the lower MS are anomalous, and we caution
against using the $(B-V)$ index for MS-fitting to the Pleiades and similarly
young open clusters.
   }

   \maketitle
%

\section{Introduction}
\label{sec:intro}

The distance to the Pleiades has been controversial ever since the Hipparcos
mission \citep{hip} measured parallaxes of individual stars in the cluster 
-- the mean distance to the cluster determined from individual star parallaxes
was found to be approximately 10\% shorter than distances previously
determined from main sequence (MS) fitting methods.
Specifically, the parallax measurements yield a distance of 118.34 pc,
corresponding to a distance modulus of $(m-M)_{0} = 5.37$ 
(\citealt{vanleeuwen} and references therein), whereas MS-fitting 
generally yields $(m-M)_{0} \approx 5.6$ (around 132 pc) or even longer 
(e.g. \citealt{mitchjohn,crawperry,mermilliod1981,nicolet,eggen,vandenbergpoll}).

After the release of the Hipparcos results, several authors attempted to 
address this discrepancy using various methods based on semi-empirical
MS-fitting techniques, and all found distance moduli in agreement with the 
earlier MS-fitting results \citep{pinsonn,soderblom,stello}.
On the other hand \citet{castellani02}, who used purely theoretical 
isochrones in their fits, were able to retrieve the Hipparcos parallax 
distance by assuming a substantially subsolar metallicity for the cluster.
Concerns were raised that most of the distance determinations in the 
literature use methods which have some model dependence which may introduce
unquantified systematic errors, hence in an earlier paper 
(\citealt*{perci2}, hereafter PSK03) we developed a purely empirical 
MS-fitting method.  This method employs a sample of 54 local field stars, 
all with precise Hipparcos parallaxes, and with homogeneous metallicity 
determinations on a scale consistent with that of the open clusters being
studied.  These field stars were used to construct a template MS which was 
fit to several open clusters, including the Hyades and the Pleiades 
(PSK03; \citealt{perci3}).
Fitting in both the $(B-V)$ and $(V-I)_{C}$ colour planes\footnote{Note 
that throughout this paper $(V-I)$ always refers to $(V-I)_{C}$, i.e. 
photometry in the Cousins system.  Some cluster data used here has been 
converted from other photometric systems, as detailed in PSK03.}, this method 
precisely reproduces the Hipparcos distance to the Hyades, yielding 
$(m-M)_{0}=3.33\pm0.06$.  Applying the same method to the Pleiades, we found 
that the $(B-V)$ fits yield $(m-M)_{0}=5.76\pm0.06$ whilst $(V-I)$ gave 
$(m-M)_{0}=5.58\pm0.04$ (corresponding to 141.9 and 130.6 pc respectively).
Not only are these results much longer than the Hipparcos parallax
distance, they are also mutually inconsistent.

The location of the MS in a colour-magnitude diagram (CMD) is, of course, 
dependent on metallicity and hence MS-fitting methods rely on matching the
metallicity of the template MS to that of the cluster.  In PSK03 we used the 
generally accepted spectroscopic metallicity for the Pleiades in our 
`standard' fits, specifically ${\rm [Fe/H]}=-0.03 \pm0.06$ (this value, plus 
error bar is taken from the catalogue of \citealt{gratton} and is based on 
the HRS determination of \citealt{boesgaardfriel}).  Since the $(B-V)$ and 
$(V-I)$ colour indices have different sensitivities to metallicity, the
discrepancy found between the $(B-V)$ and $(V-I)$ MS-fitting results suggests
that the spectroscopic metallicity is not the appropriate one to use when
applying MS-fitting to the Pleiades.  Assuming that both the $(B-V)$ and
$(V-I)$ colours are normal for some metallicity, PSK03 used colour-colour 
diagrams to demonstrate that the photometric metallicity of the Pleiades is
consistent with ${\rm [Fe/H]}=-0.4$, and hence argued that this is the
appropriate metallicity to use for MS-fits to the cluster.  Repeating the 
MS-fits using this lower metallicity, PSK03 found that consistent distances 
were obtained between the two colour planes.  Furthermore, the concordant 
distances derived at this assumed metallicity are much shorter than those 
obtained from the solar abundance fits, and are thus consistent with the 
Hipparcos parallax distance (see PSK03 for full details, and summary in 
Section~\ref{sec:pleiMSfit}, Table~\ref{tab:plei_results}). 
Hence, in PSK03 we concluded that our analysis did not support any 
mismatch between the MS-fitting and Hipparcos distances for the Pleiades
and that the widely discussed discrepancy is just an artifact due to the 
cluster's $(B-V)$ and $(V-I)$ colours (and hence, photometric metallicity), 
which are inconsistent with the spectroscopic metallicity.

However, the three most recent studies of the Pleiades, which use 
alternative
distance determination methods, once again find long distances in general 
agreement with the earlier MS-fitting results.  \citet{pan} examined data 
for Atlas (the second brightest star in the Pleiades), a wide binary which 
is resolved using optical interferometry.  Combining orbital data
with an assumed mass-luminosity relation, they determine a distance of
$133 < D < 137$ pc, with a firm lower bound of $D > 127$ pc.  The results
are slightly model-dependent, since the masses of the two components must be
taken from model isochrones -- however a 10\% uncertainty on the mass only
leads to a 3\% uncertainty on the final derived distance, due to the 
precision of the orbital parameters.  Meanwhile, \citet{munari} studied 
the eclipsing binary HD~23642.  From extensive new observations in Johnson 
$B$ and $V$ (to obtain light curves) and high resolution spectra (yielding 
radial velocities) they modelled the system and found a distance of 
132$\pm$2 pc.  These results are also slightly model dependent as the 
temperature of the primary star must be determined independently -- this is 
done by comparing photometry in many different systems (from the 
literature) to synthetic spectra.
Most recently, \citet{soderblom2004} have measured trigonometric parallaxes
for three G and K dwarfs in the Pleiades, using the Fine Guidance Sensors on
HST.  Their net parallax result of $7.43\pm0.17$ milliarcsec, corresponding to
a distance of $135\pm3$ pc, is in excellent agreement with both \citet{pan} 
and \citet{munari}. 

Hence the problem of the Pleiades distance discrepancy still exists -- whilst
empirical MS-fitting using the $(B-V)$ and $(V-I)$ colours now yields a 
distance in agreement with the Hipparcos one (by assuming the photometric 
metallicity rather than the spectroscopic one), other methods continue to 
find distances in agreement with the earlier, pre-Hipparcos, results.

It should be pointed out at this stage however that there is a known 
problem with the Pleiades, which may or may not be related to the distance 
discrepancies (we hope to clarify this point later).  This is that the lower 
main sequence stars (the K dwarfs) fall well below the position expected for 
a solar metallicity system in the $V/(B-V)$ colour-magnitude diagram.  Hence 
these stars are either underluminous, or have $(B-V)$ colours which are too 
blue for their (spectroscopic) metallicity.  As discussed in detail by 
\citet{stauffer2003} (hereafter S03), this fact has been known for many 
years, but has largely been ignored, and certainly not explained.  
S03 present new spectroscopic observations of several Pleiades K dwarfs and, 
by comparison with similar observations of K dwarfs in Praesepe, they show 
that the unusual blue colours in the Pleiades arise from anomalous spectral 
energy distributions (SEDs) -- at least for the two Pleiades K dwarfs that 
they study.  This anomaly, which S03 ascribe to rapid stellar rotation and 
`spottedness', causes the Pleiades stars to be approximately 10\% brighter in
the $B$-band than their Praesepe counterparts, whilst the $V$-band flux is
unaffected.  These spectroscopic measurements indicate that the $(B-V)$ 
colours of the Pleiades K dwarfs should be about 0.1 mag bluer in $(B-V)$ 
than the `standard' solar sequence, in agreement with the broadband 
observations. 

Comparison of the shape of the MS for the two clusters shows that in the
$V/(B-V)$ CMD the two sequences start to diverge at $M_{V} \approx 6.0$, 
in the sense that the Pleiades MS becomes increasingly bluer (or fainter) 
than the Praesepe MS towards fainter magnitudes (see S03, their Figure 4).  
Crucially, this divergence occurs right in the middle of the magnitude 
range of the field dwarfs used in our empirical MS-fitting work, which have
$5.4 \leq M_{V} \leq 7.0$.  However, similar comparisons in the $V/(V-I)$ 
and $V/(V-K)$ planes show that the two sequences have identical shapes in 
the full magnitude range of our field dwarfs.  This fact is very important
for our study, as it implies that only the $(B-V)$ index is affected by an
anomaly, and that MS-fitting using $(V-I)$ and $(V-K)$ should give reliable
(and consistent) results using the spectroscopic [Fe/H] for the cluster.  
We note here that when 
applying a MS-fitting method, for a fixed metallicity, the bluer (or more 
subluminous) the cluster MS, the longer the derived distance will be -- 
hence the PSK03 results for $(B-V)$ and $(V-I)$, at ${\rm [Fe/H]}=-0.03$, 
are consistent with the observations of S03.

With the release of the 2MASS All Sky Catalogue, we are now able to test
our empirical MS-fitting method using the infrared colours in addition to 
the optical bands.  We now know (from S03) that MS-fitting using $(B-V)$ 
colours may give spurious results for the Pleiades, therefore it is important 
to apply the method using the $(V-K)$ colours to see whether they yield
a distance which is consistent with the $(V-I)$ result, since both indices 
appear to be unaffected by any colour anomalies in the magnitude range of 
interest.  
Also, if concordant distances are obtained, we vitally need to know whether
they are consistent with the Hipparcos parallax distance, or the longer 
distance determined from other methods.  The 2MASS data also allows us to 
utilise the $(J-K)$ colours.  Whilst we have no specific information on the 
SEDs of the Pleiades K dwarfs in this part of the spectrum, the $(J-K)$ index 
should be much less sensitive to differences in metallicity than either 
$(V-I)$ or $(V-K)$, and is also much less affected by extinction.  Using the 
$K/(J-K)$ CMD also ensures that the photometry is completely homogeneous 
between cluster and field stars and thereby minimizes systematic errors.
Using $(J-K)$ and $(V-K)$ colour indices in addition to $(V-I)$ helps 
to constrain the cluster metallicity (and reddening) since the sensitivities 
are different for each colour index and consistent distances must be 
obtained from all the colour planes used if the results are valid.  
This should enable us to determine once and for all whether the discrepancy 
between the MS-fitting distance and the Hipparcos parallax distance for
the Pleiades is real. 

As in PSK03, the method can be tested on the Hyades since fits using the 
$(V-K)$ and $(J-K)$ must also be able to reproduce the Hipparcos distance 
modulus for the cluster.  We can also use M67 as a comparison cluster for the
Pleiades since it has a similar spectroscopic metallicity 
(${\rm [Fe/H]}=0.02\pm0.06$ according to \citealt{gratton}), its $(B-V)/(V-I)$
colour-colour plot is consistent with solar metallicity field dwarfs (PSK03)
and its distance is not disputed (see Sect.~\ref{sec:m67MSfit}).

The layout of the rest of the paper is as follows:  Section 2 lists the 
sources of data used in this work; Section 3 gives a brief overview of the 
empirical MS-fitting method, and presents the results of its application to 
the Hyades, M67 and the Pleiades, and Section 4 contains a discussion of the 
results and some general conclusions. 


\section{Data sources}
\label{sec:data}

The empirical MS-fitting method used here is described in detail in PSK03, to
which we refer the interested reader.  In outline, the method utilizes a 
sample of 54 local unevolved (MS) field stars, with metallicities in the
range $-0.4 \leq {\rm [Fe/H]} \leq 0.3$ for which PSK03 obtained new 
$BV(RI)_{C}$ photoelectric photometry.  All the stars in the sample have 
Hipparcos parallax measurements with errors less than 12\%, Hipparcos 
catalogue entries were also carefully checked to avoid the inclusion of any 
binaries, and metallicities were determined from their Str\"{o}mgren 
indices.  $JHK$ data were 
extracted from the 2MASS all-sky point source catalogue\footnote{hosted by 
http://irsa.ipac.caltech.edu/} using a star-by-star 
coordinate search.  Combining the 2MASS magnitudes with the parallax data 
yields the absolute magnitudes, $M_J$, $M_H$ and $M_K$, which are then 
corrected for Lutz-Kelker bias, as described in PSK03.  We note here the 2MASS
$JHK$ data have already been retrieved and utilized by \citet{sarajedini04} in 
a MS-fitting study of several open clusters, and appear in their Table 2.

2MASS data for the Hyades were retrieved for all stars listed in 
\citet{perryman} as single, definite cluster members, again using a 
coordinate search.  Data for the Pleiades were obtained in a similar manner, 
individual MS stars having been identified from the Mermilliod's WEBDA 
database\footnote{http://obswww.unige.ch/webda/webda.html} (and see PSK03) 
and known binaries removed \citep{bouvier97,raboud}.  For M67, the single
star sequence was taken from Table 5 of \citet{sandquist}, individual stars
and their coordinates were then identified by cross-correlating with the 
data of \citet{montgomery}.

Since the field star and cluster data from 2MASS are all in the same 
photometric system, for which the $K$ filter used is actually $K_{\rm S}$
($K$ short), no photometry conversions were necessary.


\section{MS-fitting using 2MASS colours}
\label{sec:MSfit}

\begin{figure}
   \centering
   \includegraphics[width=\columnwidth]{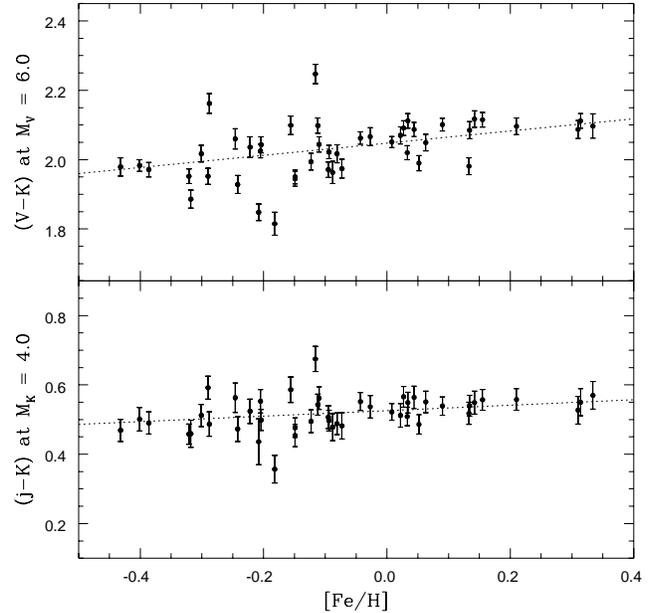}
      \caption{Metallicity dependence in $(V-K)$ and $(J-K)$.  {\bf Top panel}
               -- best fit relationship for the $(V-K)$ data, 
               $\Delta(V-K)/\Delta{\rm [Fe/H]}=0.19$;
               {\bf bottom panel} -- best fit single slope to
               the $(J-K)$ data, $\Delta(J-K)/\Delta{\rm [Fe/H]}=0.078$.
               Note that the metallicity dependence is negligible for
               ${\rm [Fe/H]} \geq 0.0$
              }
         \label{fig:fedep}
   \end{figure}

Before MS-fitting can be performed, the [Fe/H] dependence of the colour 
index used must be determined so that the field star template can be matched 
to the metallicity of the cluster -- here we use $(V-K)$ and $(J-K)$.  This 
procedure is fully described in PSK03, where it was applied to $(B-V)$ and 
$(V-I)$, and it is also utilized by \citet{sarajedini04} to determine the 
metallicity dependence of $(V-K)$.  Firstly, the slope of the relevant 
portion of the MS is estimated using the Hyades MS as a guide, then using 
this slope, each field star is `shifted' along this vector to an 
absolute magnitude of $M_{V}=6.0$.  The colour of each star at $M_{V}=6.0$ 
(i.e. $(V-K)_{M_{V}=6.0}$) is then plotted against [Fe/H] to determine the 
metallicity dependence of the colour index (see top panel of
Figure~\ref{fig:fedep}).  Using this procedure we find 
$\Delta(V-K)/\Delta{\rm [Fe/H]}=0.190$, in complete agreement with 
\citet{sarajedini04}, who find $\Delta(V-K)/\Delta{\rm [Fe/H]}=0.185$.  

When working with the $(J-K)$ colour index we will be using the $K/(J-K)$
CMD and hence the procedure is modified to find the $(J-K)$ colour at
$M_{K}=4.0$ (this roughly corresponds to $M_{V}=6.0$ for the field star
sample in that it falls near the middle of the range of magnitudes for
the sample).
The precise metallicity dependence of $(J-K)$ is harder to determine since the 
colour range of the field star sample is very small and the resultant 
$(J-K)_{M_{K}=4.0}~vs~{\rm [Fe/H]}$ plot is dominated by intrinsic 
scatter.  In fact, the total colour range of the field dwarf sample, at 
their observed colours, is only 0.25 in $(J-K)$ compared with 0.91 in $(V-K)$.

A formal fit to the field star data yields a metallicity dependence of 
$\Delta(J-K)/\Delta{\rm [Fe/H]}=0.078$
although most of this dependence comes from the stars with 
${\rm [Fe/H]} < 0.0$ (see bottom panel of Figure~\ref{fig:fedep}).  
Fitting the subsolar and supersolar metallicity 
ranges separately we find $\Delta(J-K)/\Delta{\rm [Fe/H]} = 0.086$
for ${\rm [Fe/H]} < 0.0$, whilst $\Delta(J-K)/\Delta{\rm [Fe/H]}$ is 
virtually negligible for ${\rm [Fe/H]} > 0.0$.  We note here that this 
difference in metallicity dependence for the two metallicity regimes 
is in qualitative agreement with the theoretical isochrones of 
\citet{girardi00}\footnote{available at http://pleiadi.pd.astro.it/} and 
\citet{pietrinferni}\footnote{available at 
http://www.te.astro.it/BASTI/index.php}.  However, the errors on these fits 
are of the same order 
of magnitude as the slopes, due to the relatively large scatter in
the colours and, in fact, tests using the Hyades show that the metallicity 
dependence in $(J-K)$ is actually consistent with zero (see 
section~\ref{sec:hyMSfit}).

\subsection{The Hyades}
\label{sec:hyMSfit}

The Hyades fiducials were determined by fitting a polynomial (cubic) to all 
the available single star data, reddening was assumed to be zero and the 
metallicity was taken to be ${\rm [Fe/H]} = 0.13\pm0.06$ (cluster 
metallicities are taken from \citealt{gratton}, as explained in PSK03).  
Shifting the field star sample to ${\rm [Fe/H]} = 0.13$ and fitting to the 
cluster fiducial in the $V/(V-K)$ plane gives a best-fit distance modulus of 
$(m-M)_{0} = 3.33\pm0.04$, where the 1$\sigma$ error accounts
for photometry errors for the field stars, errors on magnitudes due to their 
parallax errors, and error due to the cluster metallicity, all added in 
quadrature.

   \begin{figure}
   \centering
   \includegraphics[width=\columnwidth,height=14.0cm]{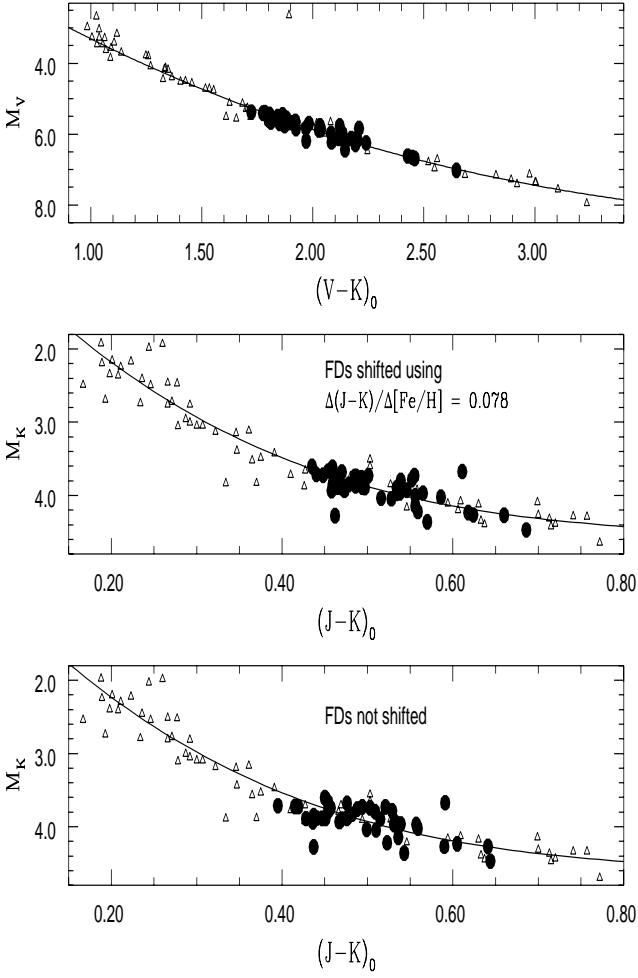}
      \caption{Fits to the Hyades MS. 
               All 3 plots show the Hyades single star data (open triangles)
               and polynomial fit to the MS line, both shifted in magnitude
               by the best-fit distance modulus, and the field star data at
               their absolute magnitudes (filled circles).               
               {\bf Top panel} -- $V/(V-K)$ CMD, using best-fit distance 
               modulus of $(m-M)_{0}=3.33$; {\bf middle and bottom panels} 
               -- $K(J-K)$ CMDs, showing shifted and unshifted field stars 
               using best-fit distance moduli of $(m-M)_{0}=3.40$ and
               $(m-M)_{0}=3.35$ respectively.
               }
         \label{fig:hy_vkjkfit}
   \end{figure}

In the $K/(J-K)$ plane, applying the maximum possible metallicity dependence
($\Delta(J-K)/\Delta{\rm [Fe/H]}=0.078$) to the field stars yields a 
best-fit distance modulus of $(m-M)_{0}=3.40\pm0.04$ which is 
marginally inconsistent with the Hipparcos parallax result of 
$(m-M)_{0}=3.33\pm0.01$.  Since the mean metallicity of the field star 
sample is ${\rm [Fe/H]} \approx -0.07$ and the Hyades is at 
${\rm [Fe/H]}=+0.13$ (i.e. near the top end of the range of metallicities 
for the field stars), a relatively small error in the estimation of 
the metallicity dependence can become significant when the field stars are 
shifted.  Hence it appears that the estimate of 
$\Delta(J-K)/\Delta{\rm [Fe/H]}=0.078$ is slightly
too high.  Repeating the MS-fit assuming no metallicity dependence in $(J-K)$
(i.e. $\Delta(J-K)/\Delta{\rm [Fe/H]}=0.0$) results in a best-fit distance 
modulus of $(m-M)_{0}=3.35\pm0.04$ (random errors only), in complete 
agreement with the Hipparcos result.  The uncertainty on the level of 
metallicity  dependence in the $(J-K)$ index induces a systematic error of
$\sim 0.06$ mag on the Hyades distance modulus since the colours of the stars 
are being `shifted' on average 0.2 dex in metallicity, with an uncertainty
of 0.078 in $\Delta (J-K)/\Delta {\rm [Fe/H]}$, and the slope of the MS in 
the $K/(J-K)$ CMD is approximately 4.  

It is important to realise however that the assumed level of metallicity 
dependence makes negligible difference to the MS-fitting results for the 
Pleiades.  This is because the metallicity of the Pleiades is very close to 
the mean of the field star sample (we assume 
${\rm [Fe/H]_{Pleiades}}=-0.03\pm0.06$ in our standard fits).  When
building the MS template, individual field stars are being shifted both ways 
in the CMD, i.e. from lower to higher and from higher to lower metallicity, 
to match the metallicity of the cluster.  Since the mean metallicity of the 
field stars is well matched to the cluster metallicity, most of the effects 
of the uncertainty on the metallicity dependence cancel out in this process. 
(see section~\ref{sec:pleiMSfit}).

\subsection{M67}
\label{sec:m67MSfit}

Fiducials for M67 were determined in the same way as for the Hyades, by
fitting a polynomial to all the available single star data.  The cluster 
metallicity was taken to be ${\rm [Fe/H]}=+0.02\pm0.06$ \citep{gratton} and 
a reddening value of $E(B-V)=0.04$ was taken from \citet{twarog97}.  
Relative extinctions and 
reddenings were calculated according to \citet{cardelli}, so that 
$A_{K}=0.114A_{V}$, $E(V-K)=2.75E(B-V)$ and $E(J-K)=0.52E(B-V)$.  After
correcting the cluster fiducial for extinction and reddening, and shifting
the field stars to the cluster metallicity, the best-fit distance modulus in
the $V/(V-K)$ plane was found to be $(m-M)_{0}=9.65\pm0.06$.  We note that this
is in complete agreement with the results of \citet{sarajedini04} who find
$(m-M)_{V}=9.74\pm0.06$ (corresponding to $(m-M)_{0}=9.62$, with 
$E(B-V)=0.04$) from a $(V-K)$ MS-fit of our field dwarf sample to the M67 
data of \citet{montgomery}. 
   \begin{figure}
   \centering
   \includegraphics[width=\columnwidth,height=14.0cm]{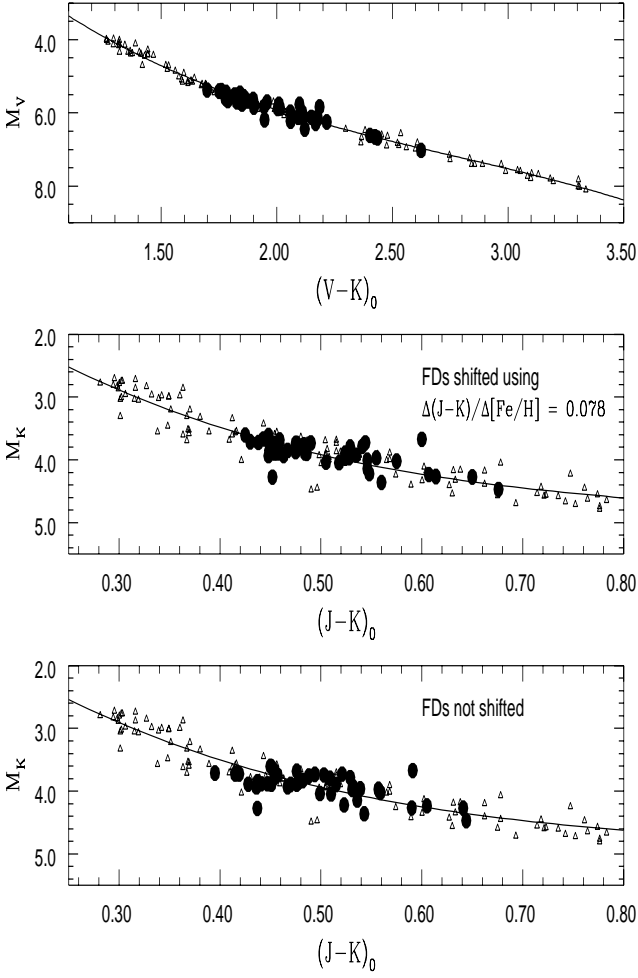}
      \caption{Fits to dereddened and extinction corrected M67 MS --
               symbols as for Figure~\ref{fig:hy_vkjkfit}.
               {\bf Top panel} -- $V/(V-K)$ CMD, using best-fit distance 
               modulus of $(m-M)_{0}=9.65$; {\bf middle and bottom panels} 
               -- $K(J-K)$ CMD, showing shifted and unshifted field stars, 
               using best-fit distance moduli of $(m-M)_{0}=9.63$ and 9.61,
               respectively.
               }
         \label{fig:m67_vkjkfit}
   \end{figure}
In the $K/(J-K)$ plane the best-fit is $(m-M)_{0}=9.63\pm0.06$ if the stars 
are shifted using $\Delta(J-K)/\Delta{\rm [Fe/H]}=0.078$, or
$(m-M)_{0}=9.61\pm0.06$ if no metallicity dependence is assumed.
The quoted 1$\sigma$
errors are as for the Hyades, but now also include the effect of an assumed 
uncertainty of 0.02 mag in $E(B-V)$.  These results are in complete
agreement with the results of MS-fitting using optical CMDs -- \citet{perci3}
find $(m-M)_{0}=9.60\pm0.09$ whilst \citet{sandquist} finds $9.60 \pm0.03$.

\subsection{The Pleiades}
\label{sec:pleiMSfit}

The Pleiades fiducials were determined in the same way as for the Hyades, by 
fitting a polynomial to all the single star data.  As in PSK03, reddenings 
for individual stars were taken from \citet{breger} (for those with no 
listing an average of $E(B-V)=0.04$ was used) and Gratton's metallicity of 
${\rm [Fe/H]}=-0.03\pm0.06$ was assumed in the standard fits.
Applying reddening and extinction corrections as before and shifting the
field stars to the cluster metallicity, the best-fit distance modulus in
$V/(V-K)$ is $(m-M)_{0}=5.67\pm0.06$, where the 1$\sigma$ error includes an
uncertainty of 0.02 mag in $E(B-V)$, as for M67.
In the $K/(J-K)$ plane, the best-fit yields $(m-M)_{0}=5.61\pm0.05$ -- it 
is important to note that in this plane, the results are the same, to within
0.01 mag, whether or not a metallicity dependence is assumed in $(J-K)$. 

   \begin{figure}
   \centering
   \includegraphics[width=\columnwidth,height=14.0cm]{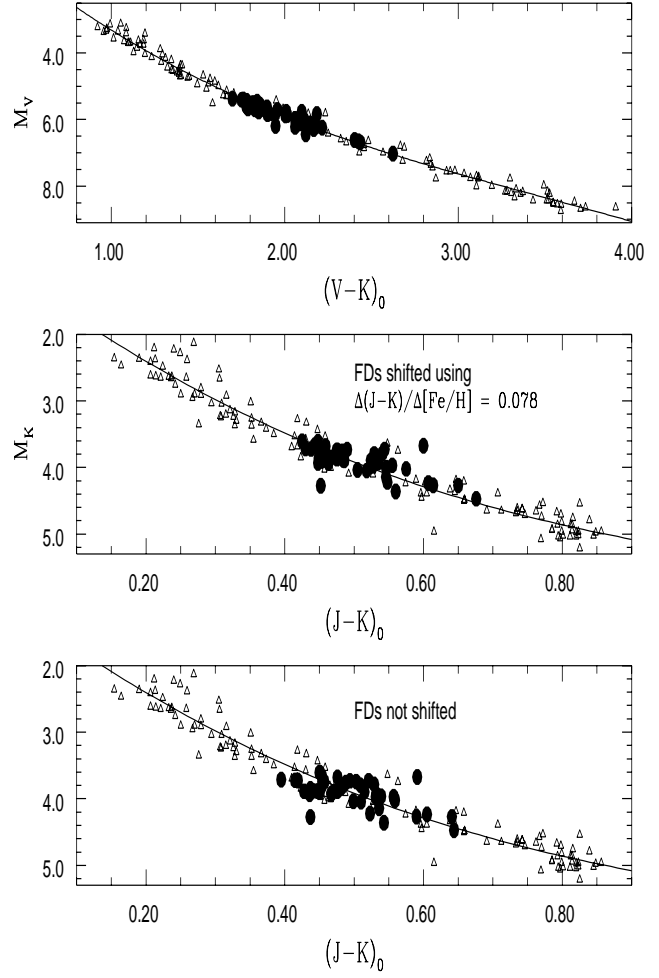}
      \caption{Fits to dereddened and extinction corrected Pleiades MS --
               symbols as for Figure~\ref{fig:hy_vkjkfit}.
               {\bf Top panel} -- $V/(V-K)$ CMD, using best-fit distance 
               modulus of $(m-M)_{0}=5.67$; {\bf middle and bottom panels} 
               -- $K(J-K)$ CMD, shifted and unshifted field stars 
               respectively, both using best-fit distance modulus
               of $(m-M)_{0}=5.61$.
               }
         \label{fig:plei_vkjkfit}
   \end{figure}

In response to the findings of S03, we returned to the $(B-V)$ data and 
tested the effect of imposing a cut at $M_{V}=6.0$, the point at which the 
$V/(B-V)$ MS appears to diverge from the `standard' sequence.  Using only 
field stars with $M_{V} < 6.0$ in the fit, the best-fit distance modulus is
$(m-M)_{0}=5.67\pm0.06$, a reduction of 0.09 mag compared to the result from
the full sample (PSK03).  Significantly, imposing the same cut in the 
$V/(V-I)$, $V/(V-K)$ and $K/(J-K)$ planes produces the same results as those 
obtained from the full sample.

We recall here that PSK03 found that by assuming 
${\rm [Fe/H]}_{Pleiades}=-0.4$, the $V/(B-V)$ and $V/(V-I)$ CMDs yielded 
distance moduli in agreement with the Hipparcos parallax result (specifically,
$(m-M)_{0}=5.46$ and 5.39 respectively, using the full field dwarf sample).  
In order to test the plausibility of this cluster abundance, we performed the 
$V/(V-K)$ and $K/(J-K)$ fits again, this time using ${\rm [Fe/H]}=-0.4$,
keeping the same reddening as for the standard fits.  
The $V/(V-K)$ fit yields $(m-M)_{0}=5.53\pm0.06$ whilst the $K/(J-K)$ fits 
yield $(m-M)_{0}=5.49\pm0.05$ for $\Delta(J-K)/\Delta{\rm [Fe/H]}=0.078$
and $(m-M)_{0}=5.61\pm0.05$ when no metallicity dependence is assumed (i.e.
the same result as for ${\rm [Fe/H]}_{Pleiades}=-0.03$).
Imposing a cut at $M_{V}=6.0$, as before, has exactly the same effect on the
results as for the ${\rm [Fe/H]}=-0.03$ fits, i.e. the $V/(B-V)$ distance
modulus is reduced by 0.09 mag, to $(m-M)_{0}=5.37$, whilst the results from 
the other colour indices remain unchanged.

All these results are summarised in Table~\ref{tab:plei_results}.

\begin{table*}
\begin{center}
\begin{tabular}{lcc} 
\hline
~~~ & $(m-M)_{0}$ & $(m-M)_{0}$ \\
\multicolumn{1}{c}{CMD} & at [Fe/H]=$-$0.03 & at [Fe/H]=$-$0.4 \\
\hline
\hline
$V/(B-V)$ (full sample) & $5.76\pm0.06$ & $5.46\pm0.06$   \\
$V/(B-V)$ ($M_{V} < 6.0$ only) & $5.67\pm0.06$ & $5.37\pm0.06$   \\
$V/(V-I)$ & $5.58\pm0.04$ & $5.39\pm0.04$   \\
$V/(V-K)$ & $5.67\pm0.06$ & $5.53\pm0.06$   \\
$K/(J-K)$ (with [Fe/H]~dependence) & $5.61\pm0.05$ & $5.49\pm0.05$ \\
$K/(J-K)$ (no [Fe/H] dependence)   & $5.61\pm0.05$ & $5.61\pm0.05$ \\
\hline
\hline
\multicolumn{1}{c}{Hipparcos result}  & ~~~~~~~~ &  ~~~~~~~~ \\
\multicolumn{1}{c}{\citep{vanleeuwen}} & \multicolumn{2}{c}{$5.37\pm0.07$} \\
\hline
\end{tabular}
\caption{Summary of MS-fitting results for the Pleiades.}
\label{tab:plei_results}
\end{center}
\end{table*}

\section{Discussion and conclusions}
\label{sec:disc}

\subsection{Summary of results}

If we accept that the K dwarfs in the Pleiades really are too blue in $(B-V)$ 
for their spectroscopic metallicity, as indicated by S03, then we should not 
expect the $(B-V)$ and $(V-I)$ MS-fitting results to be the same if the full 
MS is used (i.e going fainter than $M_{V}=6.0$).  
In fact, we should treat with caution any models 
which claim to find concordant distances in these two planes, since the
colours of the lower MS (the K dwarfs) are obviously anomalous for any 
value of [Fe/H], whether solar or otherwise.

Excluding the $(B-V)$ results using the whole field dwarf sample, the distance
moduli obtained from the `standard' fits in all the other colour planes 
(i.e. $V/(V-I)$, $V/(V-K)$ and $K/(J-K)$) are in agreement with each other 
within their 1$\sigma$ errors -- and all are discrepant by at least 
3$\sigma$ with the Hipparcos result.  The average of the $(V-I)$, $(V-K)$ 
and $(J-K)$ fits is $(m-M)_{0}=5.62\pm0.05$, corresponding to $133.1 \pm3$ pc.
Imposing a cut at $M_{V}=6.0$ and including the $(B-V)$ results, the average
of the fits in all four colour planes is $(m-M)_{0}=5.63\pm0.04$, or 133.8 pc.

Assuming ${\rm [Fe/H]}_{Pleiades}=-0.4$ (as we did in PSK03), whilst bringing 
the $(B-V)$ and $(V-I)$ results into agreement with the Hipparcos parallax
distance, does not yield consistent results in $(V-K)$ and $(J-K)$.  The
$V/(V-K)$ distance modulus is still discrepant with the Hipparcos one at
the 2$\sigma$ level, whilst the $(J-K)$ result is slightly more ambiguous
because of the uncertainty on the metallicity dependence for this colour index.
However, even using the maximum dependence, the result of 
$(m-M)_{0}=5.49\pm0.05$ is still discrepant by more than 1$\sigma$ from the 
Hipparcos distance modulus.

In summary, each of the four CMDs used here has a different slope on the MS 
and a different sensitivity to the effects of reddening and metallicity.  
We can find no combination of 
${\rm [Fe/H]}$ and $E(B-V)$ which produces distance moduli which are 
consistent across all four colour planes simultaneously when the full 
magnitude range of our field dwarf sample is used.  However, imposing a cut 
at $M_{V}=6.0$ brings all the distance moduli into agreement when assuming 
the generally accepted metallicity of ${\rm [Fe/H]}=-0.03$, and average
reddening of $E(B-V)=0.04$.  Furthermore, the average of the best-fit 
distance moduli yields a Pleiades distance of $133.8 \pm3$ pc, in complete 
agreement with the results obtained from binaries in the cluster 
\citep{pan,munari} and the most recent parallax determination from HST 
\citep{soderblom2004}.

\subsection{Some caveats}

When applying the MS-fitting method to any cluster to derive its distance,
using the local field stars as a template, we are assuming that we are 
comparing like with like -- practically, this is the only sensible assumption 
we can make.  The Pleiades is known to be anomalous in several respects --
we know it is very young ($\sim$~100~Myr) and has fast rotating stars, 
which may be affecting
the $(B-V)$ colours, whilst the field stars have unknown ages but are likely
to be much older (typically a few Gyr).

Spectroscopic metallicity determinations rely on some assumptions, one
of which is a temperature scale.  The spectroscopic determination used here
(and generally regarded as the most reliable one) is from
\citet{boesgaardfriel}, who used the temperature scale of \citet{bohmvitense}.
This temperature scale is based on $(B-V)$ colours -- i.e. from an observed
colour, a temperature is inferred.  Boesgaard \& Friel determined 
temperatures for the Pleiades stars using $(B-V)$, Str\"{o}mgren $b-y$ and 
H$\beta$ photometry, all essentially in the same (blue) portion of the optical 
spectrum, as detailed in \citet{boesgaard}.  They themselves noted that 
the $(R-I)$ data was also examined for
use in determining the temperatures but go on to say that it was ``rejected  
on the basis that all calibrations gave consistently lower temperatures than 
those obtained using the three other indices''.  We remark that this is
consistent with the observation that the late-type MS stars in the Pleiades 
have $(B-V)$ colours which are 
anomalously blue when compared to indices in other parts of the spectrum,
in that a bluer colour indicates a higher temperature.  

The problem of determining which colour index yields the most appropriate
temperature is not just confined to the Pleiades.  In a study of photospheric
abundances in active binaries, \citet{morel} state that there is a tendency 
for the $(V-R)$ and $(V-I)$ indices to yield systematically lower temperatures 
than $(B-V)$.  As was the case for the Pleiades, \citet{morel} regard the 
$(V-R)$ and $(V-I)$ colour temperatures as being ``spuriously low''.  
On the other hand, in a study of the effect of chromospheric activity on the 
mean colour of late-type stars, \citet{amado} find evidence for a blue excess 
in both the $(U-B)$ and $(B-V)$ indices for active stars when compared with
quiescent ones of the same spectral type.  In fact, \citet{amado} cautions 
against the use of the $(B-V)$ index when determining fundamental parameters 
(e.g. temperature) for late-type active (i.e. young) stars and suggests that 
near-infrared colours should be better temperature indicators for these stars.
It is beyond the scope of this work to predict what effect this may have on 
the derived spectroscopic metallicity for the Pleiades and other young 
clusters, but it should be borne in mind when assuming a metallicity for 
MS-fitting.

Reddening estimates are also generally determined from a star's colours, 
assuming that they are normal for their particular spectral type.  Most of the
\citet{breger} reddenings for the Pleiades stars are derived from broadband 
$(B-V)$ or Str\"{o}mgren $b-y$ indices, both of which are in the potentially 
anomalous part of the spectrum.  If the $B$ flux is too strong for the 
late-type (lower MS) stars, and hence the colours too blue at a fixed spectral
type, this would seem to indicate that the derived reddenings would be 
under-estimated.  This would only serve to make the Pleiades problem worse 
in that assuming a higher reddening would yield an even longer distance.
However the referee has pointed out that the Pleiades reddening (and
metallicity) measurements are in fact heavily weighted towards stars of early 
spectral types (mostly B, A and F), and so the canonical reddening and 
metallicity values may not be greatly affected by the anomaly 
displayed by the K dwarfs.

\subsection{Main conclusion}
We conclude that there are two distinct, and unrelated, issues pertaining 
to the determination of the Pleiades distance: 

\begin{itemize}
\item We believe that the Hipparcos parallax distance to the Pleiades really 
is too short by approximately 10\%, as initially suspected.
\item The $(B-V)$ colours of the lower MS (the K dwarfs) are anomalously blue
and we caution against their use for MS-fitting to the Pleiades and other 
young open clusters.
\end{itemize}

More detailed spectroscopic study is urgently needed to determine the true
[Fe/H] for the Pleiades -- bearing in mind that spectroscopic 
metallicity determinations have to assume some temperature scale, any 
anomalies must be accounted for (or at least qualitatively assessed) in this 
process.

\begin{acknowledgements}
We warmly thank the referee John Stauffer for a constructive report and some
helpful comments and suggestions.
S.M.P. would like to thank P.A.J. for financial and emotional support during
a difficult time.
This research has made extensive use of the WEBDA, HIPPARCOS and 2MASS 
databases.
\end{acknowledgements}


\end{document}